\documentclass[reprint,aps,10pt,twocolumn,floats,floatfix,prl,superscriptaddress]{revtex4-1}
\usepackage{graphicx}
\usepackage{color}

\usepackage{bm}
\usepackage{amsmath}

\begin{document}

\title{Synthetic dimensions and topological chiral currents in mesoscopic rings
}

\author{Hannah M. Price}
\affiliation{School of Physics and Astronomy, University of Birmingham, Edgbaston, Birmingham B15 2TT, United Kingdom}
\author{Tomoki Ozawa}
\affiliation{Interdisciplinary Theoretical and Mathematical Sciences Program (iTHEMS), RIKEN, Wako, Saitama 351-0198, Japan}
\author{Henning Schomerus}
\affiliation{Department of Physics, Lancaster University, Lancaster LA1 4YB, United Kingdom}

\begin{abstract}
The recently-introduced concept of ``synthetic dimensions" allows for the realization of higher-dimensional topological phenomena in lower-dimensional systems. In this work we study the complementary aspect that synthetic dimensions provide a natural route to topological states in mesoscopic hybrid devices. We demonstrate this for the current induced into a closed one-dimensional  Aharonov-Bohm ring by the interaction with a dynamic mesoscopic magnet. The quantization of the magnetic moment provides a synthetic dimension that complements the charge motion around the ring. We present a direct mapping that places the combined ring-magnet system into the class of quantum Hall models, and demonstrate that topological features, combined with the magnet's anisotropy, can lead to clear signatures in the persistent current of the single-particle ground state.
\end{abstract}

\maketitle

The recent approach of ``synthetic dimensions" has been opening up new routes to realising topological energy bands in engineered systems of atoms or photons~\cite{Celi:2014,OzawaPrice}. This idea exploits the internal degrees of freedom in a system by reinterpreting, for example, $N$ distinct internal states as a set of $N$ lattice sites along an extra synthetic spatial dimension~\cite{Boada:2012}. Similar to earlier proposals for simulating higher-dimensional models by increasing lattice connectivity~\cite{Tsomokos:2010, Jukic:2013}, this provides a way to effectively increase spatial dimensionality. This means that a system with one real spatial and one synthetic dimension can access two-dimensional topological phenomena, such as quantum Hall models with topological energy bands and corresponding robust chiral edge modes that propagate one-way around the system~\cite{Celi:2014}.

Since the first theoretical works in cold atoms~\cite{Boada:2012,Celi:2014}, there has been a rapid development of both proposals and experiments to realize different implementations of a synthetic dimension~\cite{OzawaPrice,Mancini:2015,Stuhl:2015,Gadway:2015, Meier:2016, Price:2017, Livi:2016, Kolkowitz:2017, Sundar:2018, Martin:2017, Schmidt:2015,Luo:2015,Ozawa:2016,Yuan:2016,Ozawa:2017,Cardano:2017,Yuan:2018,Lustig:2019,Peterson:2019,Dutt:2019}. However, so far, this activity has primarily focused on photonic or atomic set-ups, instead of on solid-state electronic systems. In this paper, we show that the concept of a synthetic dimension, in fact, emerges naturally also in the solid-state context of a mesoscopic ring coupled to a nanomagnet.

The induction of a current around a closed ring by a magnetic flux is an archetypical mesoscopic effect \cite{buttiker1983}. Such a current can flow even in equilibrium, where the persistent current is a property of the ground state. This bears some analogies to the dc Josephson current that flows across a weak link between two superconductors with different phases of the superconducting order parameter, which can for instance be induced by a magnetic flux through a superconducting ring. In normal rings, the current occurs as a mesoscopic effect at very low temperatures, and when measured precisely agrees with the predictions of a simple single-particle picture~\cite{bluhm2009,bleszynski2009}.

\begin{figure}[t]
  \centering
  \includegraphics[width=\columnwidth]{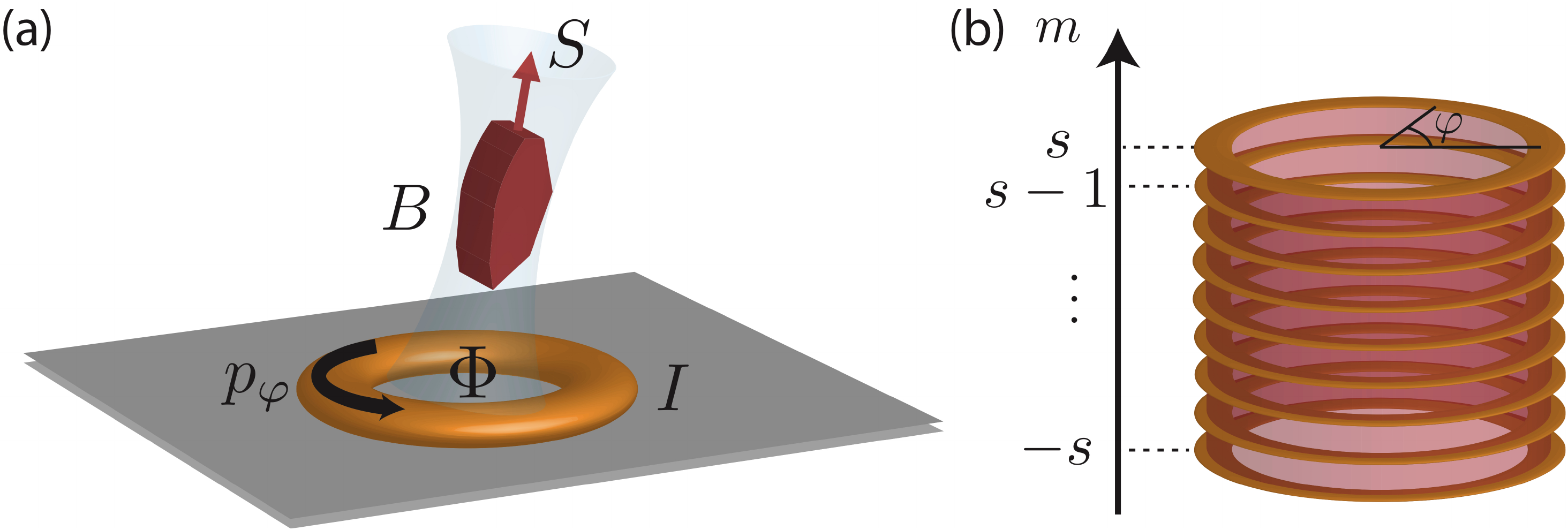}
  \caption{(a) Sketch of a mesoscopic magnet with spin $\mathbf{S}$ (spin quantum number $s\gg1$) inducing a current $I$ in a mesoscopic closed ring. The current is the response to the magnetic flux $\Phi$ induced by the field $\mathbf{B}$, which couples to the momentum $p_\varphi$ of a charge moving around the ring. (b) The situation described in (a) can be mapped to an artificial two-dimensional (cylindrical) lattice indexed by the angular coordinate $\varphi$ of the particle on the ring and the spin quantum number, $m$ of the nanomagnet. The former corresponds to a continuous and periodic (real) spatial dimension, while the latter is a discrete (synthetic) dimension with $2s+1$ sites.}\label{fig1}
\end{figure}

In this paper, we study the case where the magnetic flux through the ring is created by a dynamical mesoscopic nanomagnet. After introducing the set-up, we show that this hybrid system can be understood in a single-particle picture as a synthetic two-dimensional lattice spanned by the position of the electron around the mesoscopic ring together with the magnetic spin states, which serve as a discrete synthetic dimension. The coupling of the electron to the nanomagnet can be re-interpreted as an artificial magnetic field in this synthetic two-dimensional lattice, mapping the model to a two-dimensional quantum Hall system. As we show, the mesoscopic ring-magnet system is characterised by topological energy bands and topological ``edge" modes, corresponding to spin-polarized chiral currents around the ring. These topological modes can lead to characteristic spin-switching jumps in the ground-state persistent current. These results show that synthetic dimensions are relevant also in the solid state, providing a useful viewpoint for revealing topological effects in a setting with potential electronic applications.

{\it Setup:}
In our hybrid system, a magnetic flux passing through a mesoscopic ring is created by a nanomagnet, represented by a total spin $\mathbf{S}$ with spin quantum number $s$, as shown schematically in Fig.~\ref{fig1}(a). The total magnetic flux piercing the ring depends on the magnet's spin state, as labelled by the quantum number $m = -s, -s+1, \cdots, s$. The presence of the flux induces electrons to move around the ring, such that the system is characterised by the spin state, $m$, together with the angular coordinate, $\varphi$, of the moving (spinless) charge.

The key concept of our proposal is to map this hybrid set-up to a two-dimensional quantum Hall system on a cylinder, by re-interpreting each spin state $m$ as a different lattice site along a synthetic dimension, as sketched in Fig.~\ref{fig1}(b). As we will show, this analogy goes far. The combined magneto-electronic states form flattened-out bands with nonzero Chern numbers and chiral edge states according to the bulk-boundary correspondence. These chiral edge states have a simple interpretation in terms of a robust mesoscopic spin-orbit locking. As we furthermore show, this effect leaves direct signatures in the persistent current in the ground state of the combined system.

\textit{Hamiltonian of the hybrid system:} Before illustrating the mapping to a synthetic dimension, we introduce the Hamiltonian of the hybrid ring-magnet system [Fig.~\ref{fig1}(a)]. We shall assume that the magnetic moment of the nanomagnet is aligned with its spin so that the flux induced into the ring is proportional to the spin operator $S_z$, which at the same time is taken to align with one of the principal axes of the magnet's matrix of inertia. The spin quantum numbers $m$ introduced above then correspond to the eigenvalues of  $S_z$. In a suitable gauge, the vector potential felt by a charged particle on the ring has a component $A_\varphi \propto \rho S_z$ where $\rho$ is the ring radius. In appropriate units, the Hamiltonian for a single electron then takes the simple form
\begin{equation}
\label{eq:hamiltonian}
H=\frac{1}{2M\rho^2}\left(-i \nabla_\varphi-\gamma S_z\right)^2+\alpha S_x^2 + \beta S_y,
\end{equation}
where the kinetic-energy term includes the effect of the magnetic flux via minimal coupling with strength $\gamma$. The remaining terms describe the dynamics of the magnet as a rigid body, where $\alpha S_x^2$ determines its anisotropy while additional quadratic terms can be partially eliminated using the fact that $S_x^2+S_y^2+S_z^2$ is a c-number (thus absorbing a constant into the energy).
The term $\beta S_y$ inducing a preferred orientation is included for completeness, while additional linear terms do not further affect the robust aspects on which we concentrate here.

\textit{Mapping to a synthetic dimension:} We map the Hamiltonian (\ref{eq:hamiltonian}) to a 2D quantum Hall system, by re-interpreting the eigenstates $|m\rangle$ of $S_z$ as different lattice sites along a synthetic dimension [Fig.~\ref{fig1}(b)]. In this language, the $S_x$ and $S_y$ operators lead to ``hoppings" along the synthetic dimension, as they can be expressed by ladder operators as $S_x\! =\! (S_+ + S_-)/2$ and $S_y \!=\! (S_+ - S_-)/2i$ with
\begin{align}
S_+ |m\rangle &= \sqrt{s(s+1)-m(m+1)}|m+1\rangle \equiv 2 t_m |m+1\rangle \notag \\
S_-|m\rangle &= \sqrt{s(s+1)-m(m-1)} |m-1\rangle = 2 t_{m-1}|m-1\rangle. \notag
\end{align}
Using these relations, the Hamiltonian (\ref{eq:hamiltonian}) can be cast into a second-quantized form,
\begin{align}
	&\hat{H}
	=
	\sum_{m}\int_0^{2\pi}d\varphi\, \left[
	\frac{\left\{ \left( i \nabla_\varphi - \gamma m \right)\hat{c}^\dagger_{\varphi,m}\right\}
	\left\{ \left( -i \nabla_\varphi - \gamma m \right)\hat{c}_{\varphi,m}\right\}
}{2M \rho^2}	\right.
	\notag \\
	&+
	\alpha \left( t_m^2 + t_{m-1}^2 \right)\hat{c}^\dagger_{\varphi,m}\hat{c}_{\varphi,m}
	+ \alpha \left(  t_m t_{m+1} \hat{c}^\dagger_{\varphi,m+2}\hat{c}_{\varphi,m} + \mathrm{H.c.}\right)
	\notag \\
	&\left.+ \beta \left( -i t_m \hat{c}^\dagger_{\varphi,m+1}\hat{c}_{\varphi,m} + \mathrm{H.c.} \right)
	\right],
	\label{2ndham}
\end{align}
where $\hat{c}^\dagger_{\varphi,m} (\hat{c}_{\varphi,m})$ creates (annihilates) an excitation in the state indexed by $\varphi$ and $m$. Note that, as this is a hybrid system, these excitations physically correspond to the occupation of a composite state $|\varphi, m\rangle$, with the electron at angular position $\varphi$ on the ring and the nanomagnet in the spin state $m$.

Reinterpreting both $m$ and $\varphi$ as spatial coordinates, the Hamiltonian (\ref{2ndham}) is analogous to that of a single particle on a 2D cylinder [Fig.~\ref{fig1}(b)]. The dimension spanned by $\varphi$ is continuous and periodic ($\varphi + 2\pi = \varphi$), while that spanned by $m$ is discrete with a finite number of lattice sites, $N=2 s+1$. Under this mapping, the first line in the Hamiltonian (\ref{2ndham}) describes the kinetic energy along the $\varphi$ direction, whereas the second and third lines provide the on-site potential energy as well as hoppings (kinetic energy) along the spin ($m$) direction. Furthermore, the factor of $-\gamma m$ appearing in the kinetic energy in the $\varphi$ direction can be recognised as a vector potential in the Landau gauge, corresponding to a uniform synthetic magnetic field through the $\varphi$--$m$ plane. Thus this hybrid system mimics a particle moving on a cylinder in the presence of a uniform magnetic field.

\textit{Relationship to standard quantum Hall systems:} As we verify below, our model (\ref{2ndham}) therefore belongs to the class of 2D quantum Hall systems, with energy bands characterised by nontrivial topological Chern numbers. However, unlike most standard quantum Hall Hamiltonians, our system is continuous in one direction and discrete in the other. This is known as a ``coupled wire" configuration, as was first introduced theoretically as a tool for constructing fractional quantum Hall states~\cite{Kane:2002}. At a single-particle level, such coupled wire models have also recently become of interest due to experimental proposals for ultracold atoms~\cite{Budich:2017}, and, in the context of synthetic dimensions, for coupled optical cavities~\cite{Ozawa:2017}.

Compared to previously-studied coupled wire models, however, our system (\ref{2ndham}) has unusual nearest- and next-nearest-neighbour hoppings along the spin direction. For example, these hopping amplitudes are non-uniform, meaning that translational invariance is broken even away from the synthetic ``edges" of the system at $m = \pm s$. Such hopping anisotropy is often a feature of synthetic dimensions~\cite{Celi:2014,Price:2017}, and topological properties, such as the existence of chiral edge states, are expected to be robust provided that the anisotropy is sufficiently weak~\cite{Celi:2014,Price:2017}, as we now also demonstrate here.

\textit{Topological properties:} To calculate the bulk topological properties and anticipate the key signatures of the synthetic magnetic field, we first assume that the hopping along the spin direction is uniform, in order to make analytical arguments. Afterwards, we will revisit this assumption and verify that non-uniformity does not significantly modify the topological states. To proceed, we introduce the angular momentum quantum number $l_\varphi$ around the ring, which is constrained to take only integer values due to the periodicity of $\varphi$. Assuming uniform hopping, $t_m \approx t$, the Hamiltonian is
\begin{align}
	&\hat{H}
	\approx
	\sum_{l_\varphi, m}\left[
	\frac{\left( l_\varphi- \gamma m \right)^2}{2M \rho^2}\hat{c}^\dagger_{l_\varphi,m}\hat{c}_{l_\varphi,m} \right.
	\notag \\
	&
	\left.
	+ \alpha t^2 \hat{c}^\dagger_{l_\varphi,m+2}\hat{c}_{l_\varphi,m}
	-i \beta t \hat{c}^\dagger_{l_\varphi,m+1}\hat{c}_{l_\varphi,m} + \mathrm{H.c.}
	\right],
	\label{uniformham}
\end{align}
where $\hat{c}^\dagger_{l_\varphi, m}$ is the Fourier transform of $\hat{c}^\dagger_{\varphi,m}$ along the $\varphi$ direction, and we have omitted a constant energy offset term. This Hamiltonian does not couple operators with different $l_\varphi$. We thus have, for a given value of $l_\varphi$, a one dimensional tight-binding Hamiltonian along the spin direction, where the first term in (\ref{uniformham}) is a harmonic trapping potential centered at $m = l_\varphi/\gamma$ and the other terms are hopping terms. This is reminiscent of the ordinary Landau level problem, except that the spin direction, $m$, and the momentum, $l_\varphi$, are both discrete. Note that, in general, $l_\varphi/\gamma$ is not an integer, and thus the center of the trapping potential falls between lattice sites in spin direction. Due to the discreteness of the spin direction, the energy levels are not completely flat, although, practically, as we see below, we observe very flat energy levels for low-lying states in the bulk of the synthetic 2D system. The emergence of these flat Landau levels is a key signature of the synthetic magnetic field.

If we regard the energy levels as a function of $l_\varphi$ as energy bands, we can calculate their topological Chern numbers analytically, finding that the Chern number of every band is always equal to one regardless of the values of $\alpha$ and $\beta$~\cite{SupMat}. This behaviour agrees with previous studies of coupled wire models~\cite{Kane:2002,Budich:2017,Ozawa:2017}, where $\alpha=0$. The non-zero value of the Chern number implies that, from the bulk-boundary correspondence, between the $n$th and $(n+1)$th bulk energy levels there are $n$ chiral edge modes at each boundary $m = \pm s$.
Physically, these edge modes correspond to spin-dependent chiral currents flowing around the ring; as the largest and smallest spin states serve as the edges of the synthetic spin dimension, the edge-mode chirality means that there are currents mainly involving spin states $m = s$ flowing in one direction around the ring, and currents with $m = -s$ flowing in the opposite direction. This is the second key signature of the synthetic magnetic field.

While we arrived at these expectations assuming uniform hopping $t_m = t$, the result is more general. In the original Hamiltonian (\ref{2ndham}), the momentum $l_\varphi$ is also a good quantum number. As we introduce back non-uniformity of the hoppings, the bands as a function of $l_\varphi$ can become deformed, but as long as band gaps do not close, topological properties, such as the number of chiral edge modes which appear at $m = \pm s$, do not change. Nonetheless, the anisotropy of the full model leaves characteristic fingerprints, as we demonstrate below.

\begin{figure}[t]
  \centering
  \includegraphics[width=\columnwidth]{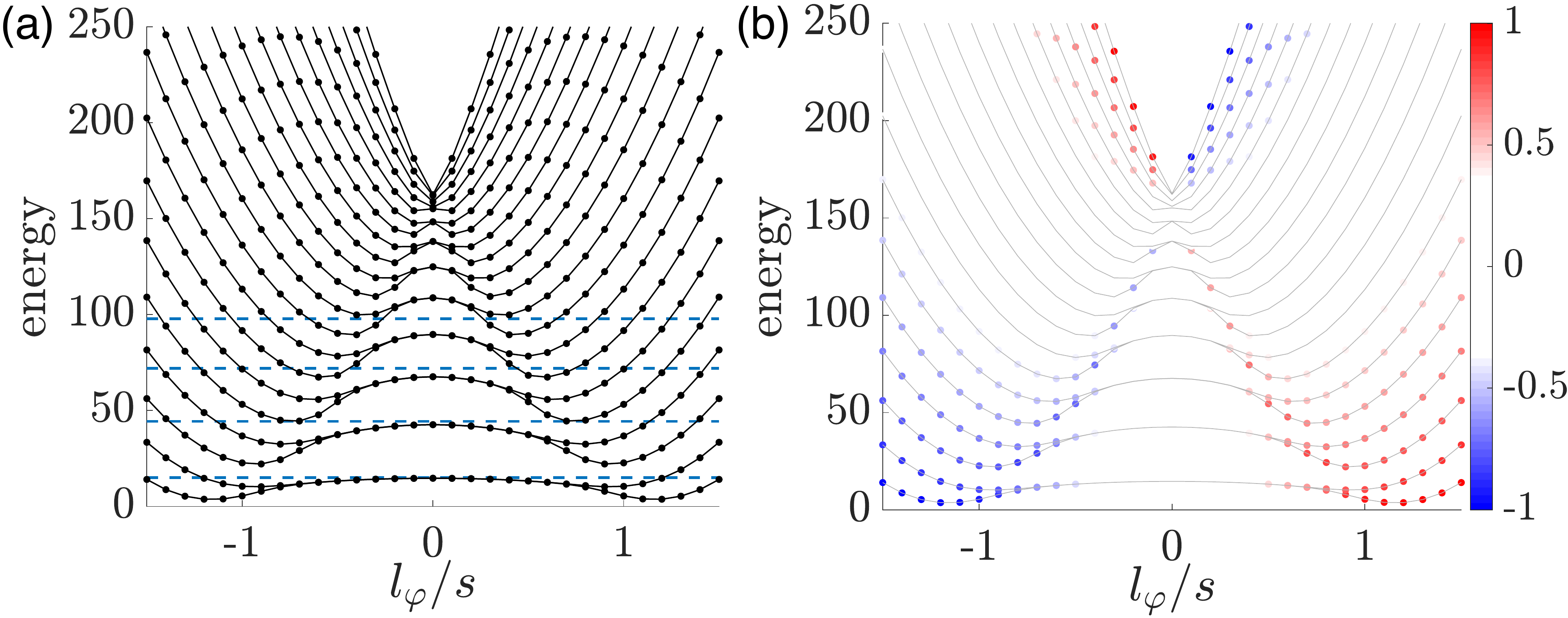}
  \caption{(a) Energy spectrum as a function of $l_\varphi / s$. We used $s = 10$, $\alpha = 1.5/(2M\rho^2)$, $\beta = 0$, and $\gamma = 1.2$. Dashed lines are ideal Landau level energies calculated with $t_m = t_0$ and $l_\varphi = 0$. Solid lines are guide to the eyes.  (b) The energy spectrum of the same system, with color representing the mean spin of the corresponding states. We observe that states corresponding to dispersive regions have spins concentrated around the edge of the synthetic dimension ($m = \pm s$), whereas low-lying nondispersive regions have spin concentrated away from the edge in the synthetic dimension. This spin dependence of the energy dispersion shows that we have chiral edge states in the synthetic two-dimensional lattice.}\label{blackcolor}
\end{figure}

\textit{Energy spectrum:} To illustrate the topological robustness of the edge states in the full anisotropic model (\ref{2ndham}) we plot the energy spectrum in Fig.~\ref{blackcolor}. Panel (a) shows the energy spectrum as a function of $l_\varphi$, normalized by the total spin $s = 10$ with $\alpha = 1.5/(2M\rho^2)$, $\beta = 0$, and $\gamma = 1.2$. The low-lying levels with small $|l_\varphi|$ show weak dispersion; these are almost-flat Landau levels. Since $\beta = 0$, states with even $m$ and odd $m$ do not couple, so that these levels are nearly two-fold degenerate. Plotted in dashed lines are ideal bulk Landau level energies calculated from Eq.~(\ref{uniformham}), assuming $t_m$ to be equal to $t_0$ and $l_\varphi = 0$. They agree well with the energy spectrum for low-lying states, verifying that the introduction of non-uniform hopping $t_m$ does not alter the basic phenomenon of bulk Landau levels.

Away from the central region, we see that Landau levels split into two and eventually move up in energy, corresponding to chiral edge states, with states concentrated around $m = s$ propagating in one direction and those concentrated around $m = -s$ propagating in the opposite direction. To clearly see the nature of these chiral edge states, we plot the spin expectation value of these eigenstates in color in Fig.~\ref{blackcolor}(b). States in red are concentrated on the edge at $m = s$, and states in blue are on the other edge at $m = -s$, showing that the propagating states are edge states in the synthetic dimension.
Since each bulk Landau level eventually moves up in energy and becomes an edge state as $|l_\varphi|$ increases, the net number of chiral edge modes propagating in the same direction is equal to the number of bands below the energy. This observation is in agreement with the fact introduced above that each Landau level has a Chern number of one. 

Including a nonzero $\beta$ splits the nearly two-fold degeneracy of the Landau levels, but otherwise does not  significantly alter the energy spectrum~[\onlinecite{SupMat}]. In particular, the bulk levels remain flat, and each bulk level still leads to one chiral edge mode. The robustness of these edge modes in the energy spectrum reflects their topological nature.

\begin{figure}[t]
  \centering
  \includegraphics[width=\columnwidth]{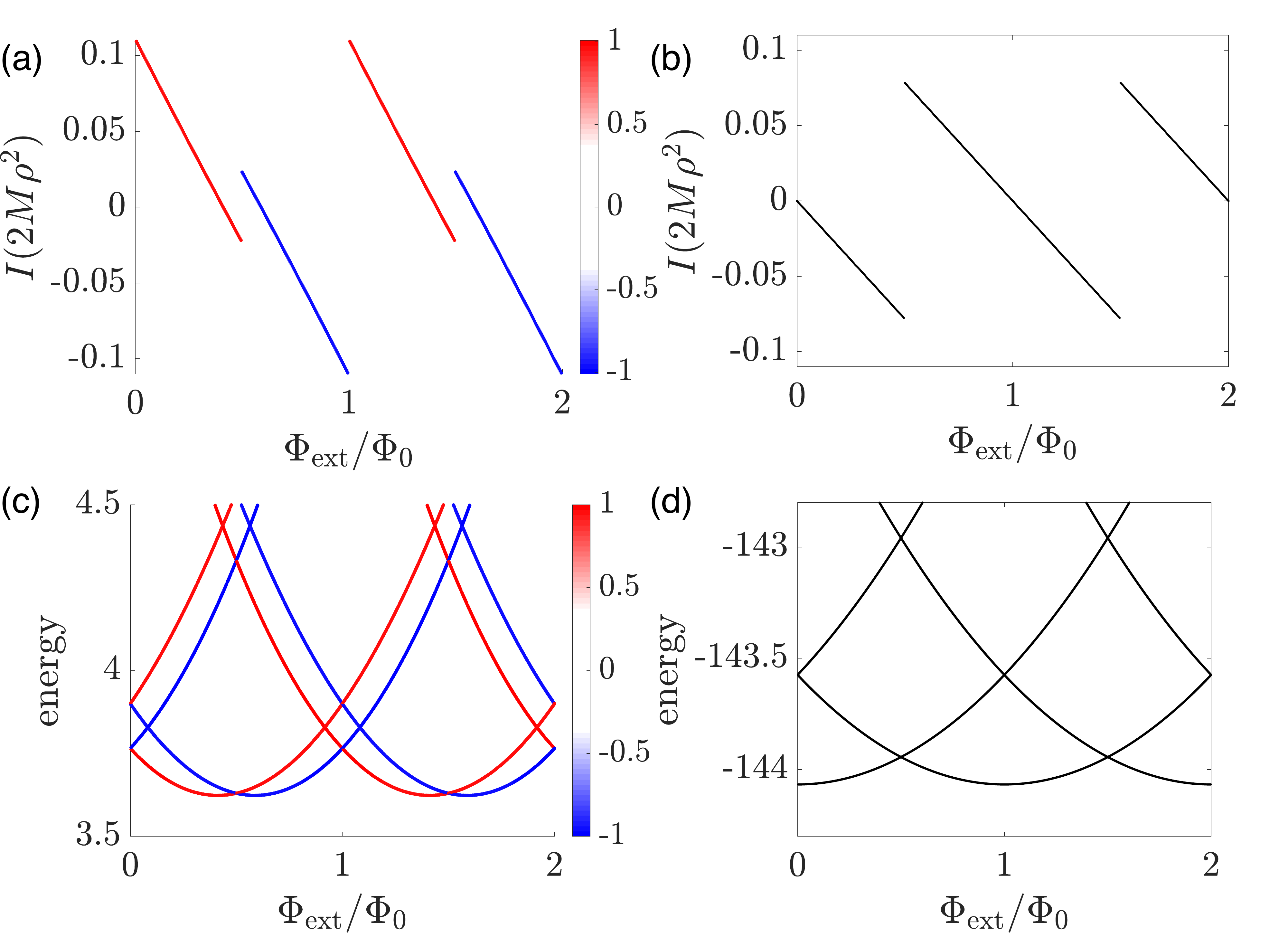}
  \caption{Persistent current  (\ref{eq:persistent}) in the single-particle ground state as a function of the externally-controlled flux $\Phi_{\rm{ext}}$, for (a) $\alpha = 1.5/(2M\rho^2)$ and (b) $\alpha = -1.5/(2M\rho^2)$, with other parameters as in Fig.~\ref{blackcolor},  where the colors represent the average spin, $\langle S_z \rangle /s$. For both (a) and (b), the persistent current jumps when there is a transition in the lowest energy state as shown in panels (c) and (d) respectively. Notably, when $\alpha>0$ in (a) and (c), the ground state is dominated by the edge modes, and the persistent current exhibits characteristic spin-switching jumps whenever the magnetic flux is increased by a half flux quantum. On the other hand, when $\alpha<0$ in (b) and (d), the ground state is localised in the middle of the synthetic dimension with $|\langle S_z \rangle / s| < 0.1$, and the persistent current only jumps if the magnetic flux is increased by a flux quantum, as is typical also in a quantum ring without a nanomagnet.}\label{persistent}
\end{figure}

\textit{Persistent currents of the single-particle ground state:} As established on general grounds above, a key hallmark of our model is the appearance of localised edge modes with respect to the synthetic dimension of spin states. Importantly, the interplay of this physics with the magnet's anisotropy also leads to clear signatures in the persistent current of the single-particle ground state of the combined system. To illustrate this, we now pierce the ring with an extra externally-controlled flux $\Phi_{\rm{ext}}$, so that Hamiltonian (\ref{eq:hamiltonian}) becomes
\begin{equation}
H=\frac{1}{2M\rho^2}\left(i \nabla_\varphi-\gamma S_z -\frac{\Phi_{\rm{ext}}}{\Phi_0}\right)^2+\alpha S_x^2 + \beta S_y, \label{eq:flux}
\end{equation}
where $\Phi_0$ is the magnetic flux quantum. The azimuthal particle velocity is then $ \dot{r}_\varphi = - (1/e)\partial H/\partial A_\varphi $, where $A_\varphi$ is the total magnetic vector potential due to both the nanomagnet and the external flux. This gives an expression for the persistent current around the ring as
\begin{eqnarray}
I = \frac{\langle \dot{r}_\varphi \rangle }{ 2\pi \rho}=  \frac{1}{2 \pi M \rho^2} \left( l_\varphi  -\gamma \langle S_z \rangle -\frac{\Phi_{\rm{ext}}}{\Phi_0} \right), \label{eq:persistent}
\end{eqnarray}
where the expectation values are taken with respect to the single-particle ground state. The persistent current is shown in Fig.~\ref{persistent}(a), for the parameters of Fig.~\ref{blackcolor}, where the colors denote the mean spin of the magnet, indicating that the ground-state persistent current is associated with large spin-polarizations. This can be understood by noting that the lowest energy states in Fig.~\ref{blackcolor} are modes localised at edges of the spin synthetic dimension. Although the existence of edge modes is a general characteristic of quantum Hall systems, their dominance in the ground state is a new feature due to the magnetic anisotropy; for $\alpha >0$, as in Fig.~\ref{persistent}(a), the anisotropy favours extremal spin values, as it generates an inverted harmonic trap $\alpha (t_m^2 +t_{m-1}^2) = \alpha [s(s+1)  -m^2]$ in the synthetic dimension (\ref{2ndham}) as well as non-uniform hopping terms. Instead, if $\alpha <0$, the ground state and hence the persistent current are localised around the middle of the synthetic dimension, as is shown in Fig.~\ref{persistent}(b).

Another clear and unusual signature of the edge modes is in the spin-switching jumps of the persistent current in Fig.~\ref{persistent}(a), occuring whenever $\Phi_{\rm{ext}}/\Phi_0 \!= \!n/2$, with $n$ being any integer; in Fig.~\ref{persistent}(b), by contrast, the persistent current only jumps when $n$ is an odd integer. Physically, these jumps correspond to parameters where the ground state becomes doubly-degenerate and then switches between different values of $l_\varphi$, as shown in Fig.~\ref{persistent}(c) and (d), where we plot the lowest energy states for panels (a) and (b) respectively.

Analytically, we can easily understand the ground-state transitions in the limit $\alpha \!=\!\beta\!=\!0$, as then the Hamiltonian~(\ref{eq:flux}) predicts possible degeneracies between states with quantum numbers $(l_\varphi, m)$ and $(-l_\varphi+ 2 \Phi_{\rm{ext}}/\Phi_0 , -m)$~\cite{SupMat}. As $l_\varphi$ is an integer, this implies transitions between states with opposite spins when $ \Phi_{\rm{ext}}/\Phi_0 = n/2$. However, if $m=0$ as in the usual quantum ring without a magnet, half of these degeneracies are not real as $(l_\varphi, 0)$ and $(-l_\varphi+ 2 \Phi_{\rm{ext}}/\Phi_0 , 0)$ can indicate the same ground state~\cite{SupMat}. Numerically, we find that this distinction persists as we turn on the other parameters as shown in  Fig.~\ref{persistent}, leading to twice as many transitions in the ground state when it is localised at the edges of the synthetic dimension as compared to the center.

\textit{Conclusion:}
In conclusion, we have demonstrated that hybrid nanomagnetic-electronic mesoscopic systems can display topological effects based on the interpretation of the nanomagnet spin as a synthetic dimension. The key signatures are flat Landau-level bands and spin-polarized chiral currents. In the presence of an additional external magnetic flux, the magnet's anisotropy leads to spin-switching behavior in the persistent current of the single-particle ground state. While we have focussed on the mesoscopic interpretation of the system it is worthwhile mentioning that the components are generic and can be realized in other ways, in particular using synthetic quantum engineering. For instance, we can  envisage quantum-optical realizations in which the large spin arises from the collective interactions of two-level systems as in a Dicke model~\cite{Garraway:2011} and the flux is induced via an artificial gauge field~\cite{Ozawa:Review}, and similar atom-optical realizations in which the described effects may serve as robust tools for quantum control~\cite{Goldman:Review,Cooper:Review}.

\begin{acknowledgments}
HMP is supported by the Royal Society via grants UF160112, RGF\textbackslash{}EA\textbackslash{}180121 and RGF\textbackslash{}R1\textbackslash{}180071. TO is supported by JSPS KAKENHI Grant Number JP18H05857, RIKEN Incentive Research Project, and the Interdisciplinary Theoretical and Mathematical Sciences Program (iTHEMS) at RIKEN. HS acknowledges funding by EPSRC via Grant No. EP/P010180/1.
 \end{acknowledgments}

\clearpage
\widetext
\begin{center}
\textbf{\large Supplemental Material for ``Synthetic dimension and topological chiral currents in mesoscopic rings"}
\end{center}
\setcounter{equation}{0}
\setcounter{figure}{0}
\setcounter{table}{0}
\setcounter{page}{1}
\makeatletter
\renewcommand{\theequation}{S\arabic{equation}}
\renewcommand{\thefigure}{S\arabic{figure}}
\renewcommand{\bibnumfmt}[1]{[S#1]}
\renewcommand{\citenumfont}[1]{S#1}

\section{Calculation of topological Chern number}

Here we calculate the topological Chern number of each band of our model. The starting Hamiltonian is Eq. (3) of the main text, 
\begin{align}
	&\hat{H}
	\approx
	\sum_{l_\varphi, m}\left[
	\frac{\left( l_\varphi- \gamma m \right)^2}{2M \rho^2}\hat{c}^\dagger_{l_\varphi,m}\hat{c}_{l_\varphi,m}
	+\alpha \left(  t^2 \hat{c}^\dagger_{l_\varphi,m+2}\hat{c}_{l_\varphi,m} + \mathrm{H.c.} \right)
	+ \beta \left( -i t \hat{c}^\dagger_{l_\varphi,m+1}\hat{c}_{l_\varphi,m} +\mathrm{H.c.} \right)
	\right],
	\label{main3}
\end{align}
in which the hopping along spin direction is assumed to be uniform ($t_m = t$). We also assume there is no edge along the $m$ direction so that we can look at the bulk properties of the Hamiltonian.

The Chern number can be calculated by first obtaining the Berry curvature and then integrating the Berry curvature over momentum space. The Berry curvature should thus be a function of momenta along two directions. One tricky aspect of calculating the Berry curvature for our model is that, although the angular momentum $l_\varphi$ is a good quantum number, the momentum along the spin direction is not a good quantum number.
This is a common problem which one encounters if one wants to calculate the Chern number of Landau levels in a continuous space. 
A way to go around this problem is to use magnetic translational symmetry instead of translational symmetry to construct momenta in two directions. This defines a magnetic unit cell, which is a unit cell in which one encloses a magnetic flux of $2\pi$. For this, we consider a segment of length $a_\varphi$ along $\varphi$ direction and $a_m$ lattice sites along $m$ direction. Since $\gamma$ is the strength of the magnetic field penetrating through the synthetic two-dimensional space, $a_\varphi$ and $a_m$ should satisfy $a_\varphi a_m \gamma = 2\pi$.

We will consider the magnetic Brillouin zone in quasimomentum space spanned by $l_\varphi$ - $p_m$, assuming that the segment spanned by $a_\phi$ - $a_m$ is a magnetic unit cell in real space, and calculate the topological properties.
We assume that $\gamma = P/Q$ where $P$ and $Q$ are mutually prime integers.
The magnetic Brillouin zone has the size of $2\pi/a_\varphi$ along $l_\varphi$ direction and $2\pi/a_m$ along $p_m$ direction.
Rewriting the kinetic energy term along $\varphi$ direction as $(l_\varphi - \gamma m)^2 = \gamma (m - l_\varphi/\gamma)^2$, we see that the Hamiltonian has the periodicity of $l_\varphi \to l_\varphi + P$. Since the magnetic Brillouin zone should be periodic with periodicity of $l_\varphi \to l_\varphi + 2\pi/a_\varphi$, we see that $2\pi/a_\varphi$ should be an integer multiple of $P$. Let us write that $2\pi/a_\varphi = PZ$ where $Z$ is an integer. Then, from $a_\varphi a_m \gamma = 2\pi$, we see that $a_m = QZ$. Therefore, $a_\varphi = 2\pi/PZ$ and $a_m = QZ$; for a given $\gamma = P/Q$, we have the freedom of choosing $Z$ when choosing a magnetic unit cell in real space.

Let us take a particular $l_\varphi$ component of the Hamiltonian (\ref{main3}) and define
\begin{align}
	\hat{H}_{l_\varphi}
	=
	\sum_{m}\left[ \frac{\left( l_\varphi - \gamma m \right)^2}{2M \rho^2}
	\hat{c}^\dagger_{l_\varphi,m} \hat{c}_{l_\varphi,m}
	+\alpha \left(  t^2 \hat{c}^\dagger_{l_\varphi,m+2}\hat{c}_{l_\varphi,m} + \mathrm{H.c.} \right)
	+ \beta \left( -i t \hat{c}^\dagger_{l_\varphi,m+1}\hat{c}_{l_\varphi,m} +\mathrm{H.c.} \right)
	\right].
\end{align}
We label the eigenstates of this Hamiltonian by $n$, where $n = 1$ is the ground state, and express the $n$-th eigenstate as $f_{l_\varphi,n}(m)$. The eigenstate of the original Hamiltonian (before Fourier transformation along $\varphi$, but with the simplification that $t_m = t$) can be written as
\begin{align}
	\psi_{l_\varphi,n}(\varphi, m) = e^{il_\varphi \varphi}f_{l_\varphi, n}(m).
\end{align}
The function $f_{l_\varphi,n}(m)$ satisfies the periodicity $f_{l_\varphi + P,n}(m + Q) = f_{l_\varphi,n}(m)$, which we use frequently in the calculation below.

The wavefunction $\psi_{l_\varphi,n}(\varphi, m)$ contains momentum along only one direction. In order to obtain momenta in two directions, we now construct a linear superposition of states with different $l_\varphi$ values: $l_\varphi$, $l_\varphi \pm 2\pi/a_\varphi$, $l_\varphi \pm 4\pi/a_\varphi$, $l_\varphi \pm 6\pi/a_\varphi$, $\ldots$, given by
\begin{align}
	\phi_{n,(l_\varphi, p_m)}(\varphi, m)
	&\equiv
	\sum_{l = -\infty}^{\infty} e^{ip_m a_m l} \psi_{l_\varphi+ 2\pi l/a_\varphi,n}(\varphi, m)
	\notag \\
	&=
	\sum_{l = -\infty}^{\infty} e^{ip_m a_m l} e^{i(l_\varphi + 2\pi l/a_\varphi) \varphi}f_{l_\varphi+ 2\pi l/a_\varphi, n}(m).
\end{align}
This is a superposition of states with the same energy, and so this state is also an energy eigenstate of the original Hamiltonian. The parameters $(l_\varphi, p_m)$ can now be regarded as quasi-momenta defined within the Brillouin zone: $-\pi/a_\varphi \le l_\varphi \le \pi/a_\varphi$ and $-\pi/a_m \le p_m \le \pi/a_m$.
From this, we can finally define the magnetic Bloch states as
\begin{align}
	u_{n,(l_\varphi, p_m)}(\varphi, m)
	&\equiv
	e^{-i(l_\varphi \varphi + p_m m)}
	\phi_{n,(l_\varphi, p_m)}(\varphi, m)
	=
	\sum_{l = -\infty}^{\infty} e^{-ip_m (m - a_m l)} e^{i \varphi 2\pi l/a_\varphi}f_{l_\varphi+ 2\pi l/a_\varphi, n}(m).
\end{align}
These magnetic Bloch states fulfill the following unit-cell periodicity:
\begin{align}
	u_{n,(l_\varphi, p_m)}(\varphi + a_\varphi, m) &= u_{n,(l_\varphi, p_m)}(\varphi, m)
	\notag \\
	u_{n,(l_\varphi, p_m)}(\varphi, m + a_m) &= e^{i \varphi 2\pi /a_\varphi} u_{n,(l_\varphi, p_m)}(\varphi, m)
	= e^{i \gamma a_m \varphi} u_{n,(l_\varphi, p_m)}(\varphi, m).
\end{align}
This is exactly the periodicity that magnetic Bloch states should obey under the Landau gauge.

Now that we have the magnetic Bloch states, we can calculate the Berry connection, Berry curvature, and the Chern number.
First, let us check the normalization of our Bloch states
\begin{align}
	&\int_0^{a_\varphi}d\varphi \sum_{m=1}^{a_m} |u_{n,(l_\varphi, p_m)}(\varphi, m)|^2
	\notag \\
	&=
	\int_0^{a_\varphi}d\varphi \sum_{m=1}^{a_m} \sum_{l,l^\prime = -\infty}^\infty
	e^{-ip_m a_m (l - l^\prime)} e^{-i \varphi 2\pi (l-l^\prime)/a_\varphi}
	f_{l_\varphi+ 2\pi l/a_\varphi, n}^*(m)
	f_{l_\varphi+ 2\pi l^\prime/a_\varphi, n}(m)
	\notag \\
	&=
	a_\varphi \sum_{m=1}^{a_m} \sum_{l = -\infty}^\infty
	\left| f_{l_\varphi+ 2\pi l/a_\varphi, n}(m)\right|^2
	\notag \\
	&=
	a_\varphi \sum_{m=-\infty}^{\infty}
	\left| f_{l_\varphi, n}(m)\right|^2.
\end{align}
In the final step, we used $\left| f_{l_\varphi+ 2\pi /a_\varphi, n}(m)\right|^2 = \left| f_{l_\varphi+ PZ, n}(m)\right|^2 = \left| f_{l_\varphi, n}(m - QZ)\right|^2 = \left| f_{l_\varphi, n}(m - a_m)\right|^2$.

The Berry connection is then, for $l_\varphi$ direction, apart from the normalization factor,
\begin{align}
	A_{l_\varphi} (l_\varphi, p_m) \propto
	&i\int_0^{a_\varphi}d\varphi \sum_{m=1}^{a_m}
	u_{n,(l_\varphi, p_m)}^*(\varphi, m) \partial_{l_\varphi} u_{n,(l_\varphi, p_m)}(\varphi, m)
	\notag \\
	&=
	i a_\varphi \sum_{m=1}^{a_m} \sum_{l = -\infty}^\infty
	f_{l_\varphi+ 2\pi l/a_\varphi, n}^*(m)
	\partial_{l_\varphi}
	f_{l_\varphi+ 2\pi l/a_\varphi, n}(m)
	=
	i a_\varphi \sum_{m=-\infty}^{\infty}
	f_{l_\varphi, n}^*(m)
	\partial_{l_\varphi}
	f_{l_\varphi, n}(m).
\end{align}
Similarly, for the $p_m$ direction,
\begin{align}
	A_{p_m} (l_\varphi, p_m) \propto
	&i\int_0^{a_\varphi}d\varphi \sum_{m=1}^{a_m}
	u_{n,(l_\varphi, p_m)}^*(\varphi, m) \partial_{p_m} u_{n,(l_\varphi, p_m)}(\varphi, m)
	\notag \\
	&=
	a_\varphi \sum_{m=1}^{a_m} \sum_{l = -\infty}^\infty
	(m-a_m l)
	|f_{l_\varphi+ 2\pi l/a_\varphi, n}(m)|^2
	=
	a_\varphi \sum_{m=-\infty}^{\infty}
	m
	|f_{l_\varphi, n}(m)|^2.
\end{align}
Therefore, we have
\begin{align}
	(A_{l_\varphi}, A_{p_m})
	=
	\frac{1}{\sum_{m^\prime=-\infty}^{\infty}
	\left| f_{l_\varphi, n}(m^\prime)\right|^2}
	\sum_{m=-\infty}^\infty
	\left(
	i
	f_{l_\varphi, n}^*(m)
	\partial_{l_\varphi}
	f_{l_\varphi, n}(m),
	m
	|f_{l_\varphi, n}(m)|^2
	\right).
\end{align}
There is no $p_m$ dependence in this Berry connection. Then, the Berry curvature is
\begin{align}
	\Omega
	=
	\partial_{l_\varphi}A_{p_m}
	=
	\partial_{l_\varphi}
	\frac{\sum_{m=-\infty}^\infty m |f_{l_\varphi, n}(m)|^2}{\sum_{m^\prime=-\infty}^{\infty}
	\left| f_{l_\varphi, n}(m^\prime)\right|^2}.
\end{align}
Finally, the Chern number is
\begin{align}
	\mathcal{C}
	&=
	\frac{1}{2\pi}
	\int_0^{2\pi/a_\varphi} dl_\varphi \int_0^{2\pi/a_m}dp_m\, \Omega
	\notag \\
	&=
	\frac{1}{a_m}
	\int_0^{2\pi/a_\varphi} dl_\varphi
	\partial_{l_\varphi}
	\frac{\sum_{m=-\infty}^\infty m |f_{l_\varphi, n}(m)|^2}{\sum_{m^\prime=-\infty}^{\infty}
	\left| f_{l_\varphi, n}(m^\prime)\right|^2}
	\notag \\
	&=
	\frac{1}{a_m}
	\left[
	\frac{\sum_{m=-\infty}^\infty m |f_{2\pi/a_\varphi, n}(m)|^2}{\sum_{m^\prime=-\infty}^{\infty}
	\left| f_{2\pi/a_\varphi, n}(m^\prime)\right|^2}
	-
	\frac{\sum_{m=-\infty}^\infty m |f_{0, n}(m)|^2}{\sum_{m^\prime=-\infty}^{\infty}
	\left| f_{0, n}(m^\prime)\right|^2}
	\right]
	\notag \\
	&=
	\frac{1}{a_m}
	\left[
	\frac{\sum_{m=-\infty}^\infty m |f_{0, n}(m-a_m)|^2}{\sum_{m^\prime=-\infty}^{\infty}
	\left| f_{0, n}(m^\prime -a_m)\right|^2}
	-
	\frac{\sum_{m=-\infty}^\infty m |f_{0, n}(m)|^2}{\sum_{m^\prime=-\infty}^{\infty}
	\left| f_{0, n}(m^\prime)\right|^2}
	\right]
	\notag \\
	&=
	\frac{1}{a_m}
	\left[
	\frac{\sum_{m=-\infty}^\infty (m + a_m) |f_{0, n}(m)|^2}{\sum_{m^\prime=-\infty}^{\infty}
	\left| f_{0, n}(m^\prime)\right|^2}
	-
	\frac{\sum_{m=-\infty}^\infty m |f_{0, n}(m)|^2}{\sum_{m^\prime=-\infty}^{\infty}
	\left| f_{0, n}(m^\prime)\right|^2}
	\right]
	\notag \\
	&=
	1.
\end{align}
Therefore, regardless of the details of the hopping along $m$, which means that regardless of the details of $\alpha$ and $\beta$, the Chern number of each band is always 1.

\section{Effect of a linear term in spin}

\begin{figure}[htbp]
  \centering
  \includegraphics[width=\columnwidth]{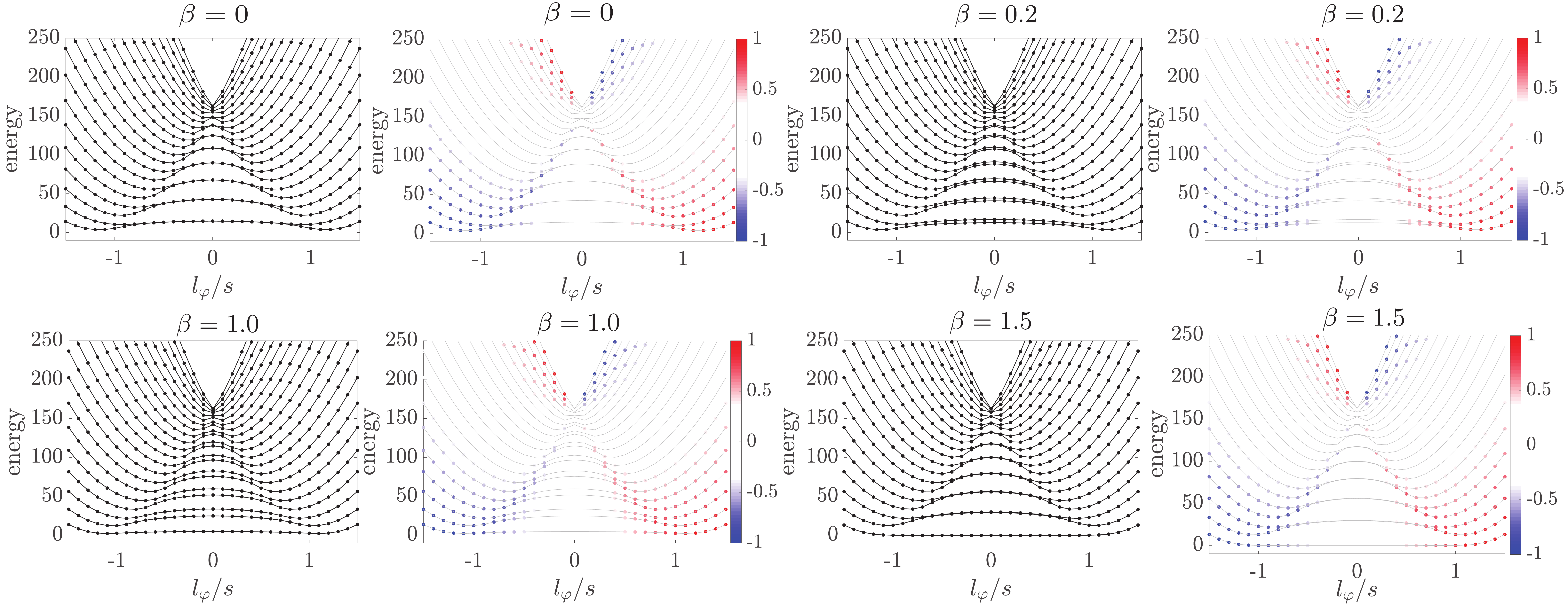}
  \caption{Energy spectrum (black-and-white) and spin expectation values (color) for various values of $\beta$, which is given in units of $1/2M\rho^2$. Other parameters are the same as in Fig. 2 of the main text, i.e., $s = 10$, $\alpha = 1.5/(2M\rho^2)$, and $\gamma = 1.2$.}\label{linearterm}
\end{figure}

Here we discuss in more detail the effect of including nonzero values of $\beta$, which is to include a linear term in spin operator in the Hamiltonian.
As discussed in the main text, when $\beta = 0$, odd sites and even sites along the spin direction are decoupled, and thus the bulk spectrum is doubly degenerate. When nonzero $\beta$ is present, this degeneracy is lifted. In Fig.~\ref{linearterm}, we plot the energy spectra and their spin expectation values for various values of $\beta$, with all the other parameters fixed to a common value. For $\beta = 0$, the spectrum is the same as the one plotted in Fig. 2 of the main text. We can see that as $\beta$ is increased from zero, the double degeneracy present for $\beta = 0$ is gradually lifted by an almost-rigid shift. When $\beta = 1.5/2M\rho^2$, the lifted levels become again almost degenerate with the higher levels, with the lowest level remaining nondegenerate. An important feature here is that, apart from this energy shift, the main features, such as the flatness of the  Landau levels in the bulk and the spin-polarization of the chiral edge states, are unchanged. This robustness of the topological features is a consequence of the result of the previous section that the Chern number of each band is one regardless of the values of $\alpha$ and $\beta$.

\section{Ground-state transitions}

In this section, we provide more detail concerning the ground-state transitions for the persistent current shown in the main text. As mentioned there, we can gain analytical insight into the transitions by starting from the limit $\alpha \!=\!\beta\!=\!0$, in which the Hamiltonian is simply
\begin{equation}
H=\frac{1}{2M\rho^2}\left(l_\varphi-\gamma m -\frac{\Phi_{\rm{ext}}}{\Phi_0}\right)^2.  \label{eq:flux}
\end{equation}
As this Hamiltonian is diagonal with respect to both the angular momentum and the magnet polarisation, each eigenstate can be labeled by the two integer quantum numbers $(l_\varphi, m)$. It can then be immediately seen from inspection of Eq.~\eqref{eq:flux} that an eigenstate labelled by $(l_\varphi, m)$ can be degenerate with eigenstates labelled by: (i) $(-l_\varphi+ 2 \Phi_{\rm{ext}}/\Phi_0 , -m)$, (ii) $(-l_\varphi+ 2 \Phi_{\rm{ext}}/\Phi_0 + 2 \gamma m , m)$, and (iii) $(-l_\varphi+ 2 \gamma m ,- m)$.

Remembering that the angular momentum quantum number must be an integer then leads to the  conditions that respectively
(i) ${ \Phi_{\rm{ext}}}/{\Phi_0}  = {n}/{2} $,
(ii) ${ \Phi_{\rm{ext}}}/{\Phi_0}  +  \gamma m = {n}/{2}$,
(iii) $ \gamma m = {n}/{2} $,
where $n$ is any integer. As can be seen, the conditions (ii) and (iii) are not generic as they require fine-tuning of parameters to ensure that the spin quantum number $m$ is an integer in the range $(-s,...,s)$ as required, and hence the occurrence of these degeneracies is highly sensitive to the value of $\gamma$. Hereafter, we therefore focus on (i) which is generic and only depends on the external flux.

As we are interested in the ground state, we can ask what value of the angular momentum, $l_\varphi$, in (\ref{eq:flux}) will lead to the lowest energy for a given $m$, finding that
\begin{eqnarray}
&l_\varphi = \frac{n}{2} + \lfloor \gamma m \rfloor \qquad & \text{for ${ \Phi_{\rm{ext}}}/{\Phi_0}  = {n}/{2} $ when $n$ is even}, \nonumber\\
&l_\varphi =  \frac{n-1}{2} + \lfloor \gamma m + \frac{1}{2} \rfloor \qquad & \text{for ${ \Phi_{\rm{ext}}}/{\Phi_0}  = {n}/{2} $ when $n$ is odd.}
\end{eqnarray}
This suggests that the lower-energy states will include degeneracies between states labelled by
\begin{eqnarray}
&(\frac{n}{2} + \lfloor \gamma m \rfloor, m )  \text{\qquad  and \qquad }  (\frac{n}{2} - \lfloor \gamma m \rfloor , -m) &  \qquad  \text{for ${ \Phi_{\rm{ext}}}/{\Phi_0}  = {n}/{2} $ when $n$ is even}, \nonumber\\
&(\frac{n-1}{2} + \lfloor \gamma m \rfloor, m )  \text{\qquad and \qquad  }  (\frac{n+1}{2} - \lfloor \gamma m \rfloor , -m)  &\qquad  \text{for ${ \Phi_{\rm{ext}}}/{\Phi_0}  = {n}/{2} $ when $n$ is odd.} \label{eq:transitions}
\end{eqnarray}
Importantly, we note that for $m=0$, the former trivialises when $n$ is even as these are in fact the same state; this is the well-known result that the persistent current of a single particle in a quantum ring jumps only at values of ${ \Phi_{\rm{ext}}}/{\Phi_0}  = {n}/{2} $, where $n$ is an odd integer.

So far, the above argument was based on assuming a given spin, $m$. However, the ground state will be the lowest energy state over all allowed values of this spin. From the Hamiltonian (\ref{eq:flux}) it can be seen immediately that $m=0$ will have zero energy whenever ${ \Phi_{\rm{ext}}}/{\Phi_0}  = {n}/{2} $, with $n$ being even. As the energy from Eq.~\ref{eq:flux} is always positive, there can be no states that are lower than zero energy, and so $m=0$ will always contribute to the ground state. (Other spin states will also be at zero energy if $ \gamma m$ is an integer, and then the ground state is a superposition of states.)

To go beyond this, we need to re-introduce terms such as those relating to the anisotropy and preferred orientation of the magnet. As we see from the full Hamiltonian presented in the main text, these additional terms are diagonal in the angular momentum and so $l_\varphi$ remains a good quantum number. In general, the eigenstates and eigenspectrum of this Hamiltonian are found numerically, as has been presented in the main text. Then, as shown in the main text, the transitions predicted by Eq.~\eqref{eq:transitions} can persist for the full Hamiltonian, leading to the spin-switching jumps observed whenever ${ \Phi_{\rm{ext}}}/{\Phi_0}  = {n}/{2} $, with $n$ being any integer.

This is possible because, for positive $\alpha$, the ground state is no longer necessarily associated with $m=0$, but can be localised at the edges of the system. This can be intuitively understood by noting that $\alpha$ contributes both non-uniform hopping terms as well as an inverted harmonic trapping potential, $\alpha (t_m^2 +t_{m-1}^2) = \alpha (s(s+1)  -m^2)$ along the synthetic dimension, which, in particular, favours sites at the edges of the synthetic dimension.

Inspired by this, we consider only the effects of this inverted harmonic trap, so that we still have $m$ as a good quantum number. For this simplified situation, the ground-state, e.g. at ${ \Phi_{\rm{ext}}}/{\Phi_0}  =n/2$, will be determined by a  Hamiltonian
\begin{equation}
H_{\text{simplified}}=\frac{1}{2M\rho^2}\left( \lfloor \gamma m \rfloor-\gamma m \right)^2 - \alpha m^2, \label{eq:2}
\end{equation}
where we have neglected an overall energy shift, and used the results obtained in Eq.~\eqref{eq:transitions}. From this we can anticipate that the ground state will shifted from $m=0$ as long as there is at least one non-zero allowed spin value, $m_0$, for which
$ \alpha m_0^2 > \frac{1}{2M\rho^2}\left( \lfloor \gamma m_0 \rfloor-\gamma m_0 \right)^2$. This is generically possible provided that $\alpha 2M\rho^2$, $\gamma$ and $s$ are large enough.

As a final comment, we also observe numerically that increasing $\beta$ serves to flatten out the dispersion, as shown above. While this is an advantage for drawing the analogy with Landau levels, this may be a disadvantage if we want to isolate only the edge modes in the ground state. Nevertheless, the predicted spin-switching jumps in the persistent current should be observable provided that $\beta$ is sufficiently small.


\begin{thebibliography}{99}%


\bibitem{Celi:2014} A. Celi, P. Massignan, J. Ruseckas, N. Goldman, I. B. Spielman, G. Juzeliūnas, and M. Lewenstein,
\emph{Synthetic gauge fields in synthetic dimensions}, Phys. Rev. Lett. {\bf 112}, 043001 (2014).

\bibitem{OzawaPrice} T. Ozawa and H. M. Price, \emph{Topological quantum matter in synthetic dimensions}, Nat. Rev. Phys. {\bf 1}, 349 (2019).

\bibitem{Boada:2012} O. Boada, A. Celi, J. I. Latorre, and M. Lewenstein,
\emph{Quantum simulation of an extra dimension}, Phys. Rev. Lett. {\bf 108}, 133001 (2012).

\bibitem{Tsomokos:2010} D. I. Tsomokos, S. Ashhab, and F. Nori,
\emph{Using superconducting qubit circuits to engineer exotic lattice systems}, Phys. Rev. A {\bf 82}, 052311 (2010).

\bibitem{Jukic:2013} D. Juki\'c and H. Buljan,
\emph{Four-dimensional photonic lattices and discrete tesseract solitons}, Phys. Rev. A {\bf 87}, 013814 (2013).

\bibitem{Mancini:2015} M. Mancini, G. Pagano, G. Cappellini, L. Livi, M. Rider, J. Catani, C. Sias, P. Zoller, M. Inguscio, M. Dalmonte, and L. Fallani,
\emph{Observation of chiral edge states with neutral fermions in synthetic Hall ribbons}, Science {\bf 349}, 1510 (2015).

\bibitem{Stuhl:2015} B. K. Stuhl, H. I. Lu, L. M. Aycock, D. Genkina, and I. B. Spielman,
\emph{Visualizing edge states with an atomic Bose gas in the quantum Hall regime}, Science {\bf 349}, 1514 (2015).

 \bibitem{Gadway:2015} B. Gadway,
 \emph{Atom-optics approach to studying transport phenomena}, Phys. Rev. A {\bf 92}, 043606 (2015).

 \bibitem{Meier:2016} E. J. Meier, F. A. An and B. Gadway,
  \emph{Atom-optics simulator of lattice transport phenomena}, Phys. Rev. A {\bf 93}, 051602 (2016).

\bibitem{Price:2017} H. M. Price, T. Ozawa, and N. Goldman,
\emph{Synthetic dimensions for cold atoms from shaking a harmonic trap},
Phys. Rev. A {\bf 95}, 023607 (2017).

\bibitem{Livi:2016} L. F. Livi, G. Cappellini, M. Diem, L. Franchi, C. Clivati, M. Frittelli, F. Levi, D. Calonico, J. Catani, M. Inguscio and L. Fallani,
 \emph{Synthetic dimensions and spin-orbit coupling with an optical clock transition}, Phys. Rev. Lett. {\bf 11} 7, 220401 (2016).

\bibitem{Kolkowitz:2017} S. Kolkowitz, S. L. Bromley, T. Bothwell, M. L. Wall, G. E. Marti, A. P. Koller, X. Zhang, A. M. Rey and J. Ye,
 \emph{SpinÐorbit-coupled fermions in  an optical lattice clock}, Nature {\bf 542} 66 (2017).

\bibitem{Sundar:2018} B. Sundar, B. Gadway and K. R. Hazzard,
 \emph{Synthetic dimensions in ultracold polar molecules}, Sci. Rep. {\bf 8} 3422 (2018).

 \bibitem{Martin:2017} I. Martin, G. Refael and B. Halperin,
 \emph{Topological frequency conversion in strongly driven quantum systems}, Phys. Rev. X {\bf 7}, 041008 (2017).

\bibitem{Schmidt:2015} M. Schmidt, S. Kessler, V. Peano, O. Painter, and F. Marquardt,
\emph{Optomechanical creation of magnetic fields for photons on a lattice}, Optica {\bf 2}, 635 (2015).

\bibitem{Luo:2015} X.-W. Luo, J.-S. Xu, G.-C. Guo, X. Zhou, C.-F. Li, and Z.-W. Zhou,
\emph{Quantum simulation of 2D topological physics in a 1D array of optical cavities}, Nat. Commun. {\bf 6}, 7704 (2015).

\bibitem{Ozawa:2016} T. Ozawa, H. M. Price, N. Goldman, O. Zilberberg, and I. Carusotto,
\emph{Synthetic dimensions in integrated photonics: From optical isolation to four-dimensional quantum Hall physics} Phys. Rev. A {\bf 93}, 043827 (2016).

\bibitem{Yuan:2016} L. Yuan, Y. Shi, and S. Fan,
\emph{Photonic gauge potential in a system with a synthetic frequency dimension}, Opt. Lett. {\bf 41}, 741 (2016).

\bibitem{Ozawa:2017} T. Ozawa and I. Carusotto,
\emph{Synthetic dimensions with magnetic fields and local interactions in photonic lattices},
Phys. Rev. Lett. {\bf 118}, 013601 (2017).

\bibitem{Cardano:2017} F. Cardano, A. D'Errico, A. Dauphin, M. Maffei, B. Piccirillo, C. de Lisio, G. De Filippis, V. Cataudella, E. Santamato, L. Marrucci, M. Lewenstein and P. Massignan,
\emph{Detection of Zak phases and topological invariants in a chiral quantum walk of twisted photons}. Nat. Commun. {\bf 8}, 15516 (2017).

\bibitem{Yuan:2018} L. Yuan, Q. Lin, M. Xiao, and S. Fan,
\emph{Synthetic dimension in photonics}, Optica {\bf 5}, 1396 (2018).

\bibitem{Lustig:2019} E. Lustig, S. Weimann, Y. Plotnik, Y. Lumer, M. A. Bandres, A. Szameit, and M. Segev,
\emph{Photonic topological insulator in synthetic dimensions}, Nature \textbf{567}, 356 (2019).

\bibitem{Peterson:2019} C. W. Peterson, W. A. Benalcazar, M. Lin, T. L. Hughes, and G. Bahl, \emph{Strong nonreciprocity in modulated resonator chains through synthetic electric and magnetic fields}, arXiv:1903.07408.

\bibitem{Dutt:2019} A. Dutt, M. Minkov, Q. Lin, L. Yuan, D. A. B. Miller, and S. Fan, \emph{Experimental band structure spectroscopy along a synthetic dimension}, arXiv:1903.07842.

\bibitem{buttiker1983}
M. B{\"u}ttiker, Y. Imry, and R. Landauer,
\emph{Josephson behavior in small normal one-dimensional rings},
Phys. Lett. A \textbf{96}, 365--367 (1983).

\bibitem{bluhm2009}
H. Bluhm, N. C. Koshnick, J. A. Bert, M. E. Huber, and K. A. Moler,
\emph{Persistent Currents in Normal Metal Rings},
Phys. Rev. Lett. \textbf{102}, 136802 (2009).

\bibitem{bleszynski2009}
A. C. Bleszynski-Jayich, W. E. Shanks, B. Peaudecerf, E. Ginossar, F. von Oppen, L. Glazman, and J. G. E. Harris,
\emph{Persistent Currents in Normal Metal Rings},
Science \textbf{326}, 272--275  (2009).

\bibitem{Kane:2002} C. L. Kane, R. Mukhopadhyay, and T. C. Lubensky, {\it Fractional quantum Hall effect in an array of quantum wires}, Phys. Rev. Lett. {\bf 88}, 036401 (2002).

\bibitem{Budich:2017} J. C. Budich, A. Elben, M. \L\c{a}cki, A. Sterdyniak, M. A. Baranov, and P. Zoller, {\it Coupled atomic wires in a synthetic magnetic field}, Phys. Rev. A {\bf 95}, 043632 (2017).

\bibitem{Garraway:2011} B. M. Garraway, \emph{The Dicke model in quantum optics: Dicke
model revisited,} Phil. Trans. R. Soc. A {\bf 369}, 1137-1155 (2011).

\bibitem{Ozawa:Review} T. Ozawa, H. M. Price, A. Amo, N. Goldman, M. Hafezi, L. Lu, M. C. Rechtsman, D. Schuster, J. Simon, O. Zilberberg, and I. Carusotto, \emph{Topological photonics,} Rev. Mod. Phys. {\bf 91}, 015006 (2019).

\bibitem{Goldman:Review} N. Goldman, G. Juzeli\={u}nas, P. \"Ohberg, and I. B. Spielman, \emph{Light-induced gauge fields for ultracold atoms,} Rep. Prog. Phys. {\bf 77}, 126401 (2014).

\bibitem{Cooper:Review} N. R. Cooper, J. Dalibard, and I. B. Spielman, \emph{Topological bands for ultracold atoms,} Rev. Mod. Phys. {\bf 91}, 015005 (2019).

\bibitem{SupMat}
See supplemental material for detailed discussions of the Chern number, preferred spin orientations, and ground state transitions.

\end{thebibliography}
\end{document}